\documentclass[10pt]{JHEP3}
\usepackage{amsmath}
\usepackage{verbatim}
\usepackage{graphicx}

\newcommand{\e}{\epsilon}

\newcommand{\tL}{\tilde{L}}
\newcommand{\tM}{\tilde{M}}
\newcommand{\tJ}{\tilde{J}}
\renewcommand{\L}{{\mathcal{L}}}
\newcommand{\bL}{\bar{{\mathcal{L}}}}
\newcommand{\z}{{\bar z}}

\renewcommand\O{{\mathcal{O}}}
\newcommand{\be}[1]{ \begin{equation}\label{#1} }
\newcommand{\ee}{\end{equation}}
\newcommand{\ben}[1]{\begin{eqnarray}\label{#1} }
\newcommand{\een}{\end{eqnarray}}
\newcommand{\eq}[1]{(\ref{#1})}
\newcommand{\p}{\partial}
\newcommand{\D}{\Delta}
\newcommand{\tD}{\tilde{\Delta}}
\newcommand{\tx}{\tilde{\xi}}
\newcommand{\w}{\omega}
\newcommand{\bw}{\bar{\omega}}
\newcommand{\refb}[1]{(\ref{#1})}
\newcommand{\<}{\langle}
\renewcommand{\>}{\rangle}
\renewcommand{\(}{\left(}
\renewcommand{\)}{\right)}

\title{BMS/GCA Redux \\ \vspace{0.3cm} \large{Towards Flatspace Holography from Non-Relativistic Symmetries.}}
\author{
Arjun Bagchi$^1$, Reza Fareghbal$^2$\\
$\;$ $\,$$^1$ School of Mathematics, \\
$\;$ $\,$ University of Edinburgh \\
$\;$ $\,$ Kings Buildings, Edinburgh EH9 3JZ\\
$\;$ $\,$ United Kingdom \\
$\;$ $\,$$^2$ School of Particles and Accelerators, \\ 
$\;$ $\,$ Institute for Research in Fundamental Sciences (IPM),  \\
$\;$ $\,$ P.O. Box 19395-5531, Tehran, Iran

$\;$\email{arjun.bagchi@ed.ac.uk, fareghbal@theory.ipm.ac.ir}
}

\abstract{The asymptotic group of symmetries at null infinity of flat spacetimes in three and four dimensions is the infinite dimensional Bondi-Metzner-Sachs (BMS) group. This has recently been shown to be isomorphic to non-relativistic conformal algebras in one lower dimension, the Galilean Conformal Algebra (GCA) in 2d and a closely related non-relativistic algebra in 3d \cite{Bagchi:2010zz}. We provide a better understanding of this surprising connection by providing a spacetime interpretation in terms of a novel contraction. The 2d GCA, obtained from a linear combination of two copies of the Virasoro algebra, is generically non-unitary. The unitary subsector previously constructed had trivial correlation functions. We consider a representation obtained from a different linear combination of the Virasoros, which is relevant to the relation with the BMS algebra in three dimensions. This is realised by a new space-time contraction of the parent algebra. We show that this representation has a unitary sub-sector with interesting correlation functions. We discuss implications for the BMS/GCA correspondence and show that the flat space limit actually induces precisely this contraction on the boundary conformal field theory. We also discuss aspects of asymptotic symmetries and the consequences of this contraction in higher dimensions.}

\preprint{EMPG-12-04, NI-12014}

\begin{document}

\baselineskip 3.5ex

\section{Introduction}

\subsection{Flat Space Holography}

The holographic gauge-gravity dualities have paved the way for much of our present understanding of quantum gravity. These correspondences relate gravitational theories or string theories to quantum field theories in lower dimensions and by performing calculations in the non-gravitational theories one can probe many mysteries that have plagued quantum gravity. But ever since the advent of the AdS/CFT correspondence \cite{Maldacena:1997re}, the spectacular success of the holographic principle in Anti de Sitter space has been in sharp contrast to the difficulties in formulating the flat space version of holography \cite{Susskind:1998vk}. The physical relevance of cosmologies and Minkowski spacetimes makes investigation of flat holography one of the important problems in string theory today.

Among the different approaches to flatspace holography, a recent one has been to consider the Bondi-Metzner-Sachs (BMS) group. In the absence of gravity, the isometry group of the spacetime is the well-known Poincare group, the semidirect product of translations and Lorentz transformations. In three and four dimensions with gravity turned on, the situation changes drastically. At null infinity in four-dimensional asymptotically flat metrics, the isometry group of the background metric is enhanced to an infinite dimensional asymptotic symmetry group. This is the BMS group \cite{BMS}, consisting  of the semidirect product of the global conformal group of the unit 2-sphere and the infinite dimensional ÔÔsupertranslationsÕÕ. There is a further enhancement to two copies of the centre-less Virasoro or the Witt algebra times the supertranslations if one does not require the transformations generated to be globally well defined \cite{bar-comp, Barnich4d}. In three dimensions the BMS group is again infinite dimensional and now has the global conformal group of $S^1$ along with the supertranslations. There is a similar enhancement to a Virasoro algebra times supertranslations even here \cite{bar-comp}. These observation lead to a recent proposal of a holographic relation between flat 3 and 4 dimensional spaces and relativistic CFTs in 1 and 2 dimensions \cite{Barnich:bmscft1,Barnich:2011ct}.

In \cite{Bagchi:2010zz}, a radically different prescription for studying flat-space holography was proposed. We proposed that instead of a relativistic 1d and 2d CFT, the holographic duality should be between 2d and 3d {\em{non-relativistic}} CFTs on the one hand and the flat 3d and 4d spacetimes on the other hand. The 3d BMS algebra was shown to be isomorphic to the non-relativistic 2d Galilean Conformal Algebra and the 4d BMS algebra to a closely related non-relativistic algebra called the semi-GCA in three dimensions. This matching of symmetries provided a first step to an intriguing connection, fancifully christened the BMS/GCA correspondence, between asymptotically flat spacetimes and non-relativistic CFTs in one lower dimension.

Various aspects of this BMS/GCA correspondence remain to be explained, the foremost of which is the spacetime interpretation of the symmetries on both sides which would lead to a starting point of a holographic dictionary. Our endeavour in this paper is to provide the initial steps in this direction and we shall primarily focus here on the BMS$_3$/GCA$_2$ case.

\subsection{Galilean Conformal Algebra}

In order to provide a spacetime interpretation of the BMS/GCA correspondence, we would concentrate on the representations of the non-relativistic algebra in two dimensions. In general dimensions, the GCA is obtained by the non-relativistic contraction of the relativistic conformal algebra and one finds it to be infinite dimensional for all dimensions of spacetime \cite{Bagchi:2009my}. In other words, the finite dimensional algebra obtained by contraction was found to have a natural embedding in an infinite-dimensional algebra (also called GCA). A natural question to ask is what happens in two dimensions where the relativistic algebra is also infinite dimensional. It was found in \cite{GCA2d} that indeed a simple relation exists between these infinite dimensional relativistic and non-relativistic algebras. One can obtain the GCA by taking a simple linear combinations of the two copies of the Virasoro algebra and taking the non-relativistic limit.

The aforementioned mapping from the relativistic Virasoro to the GCA in 2d, although very systematic and useful, has certain inherent problems. Due to the strange nature of the limit (one needs to scale the relativistic conformal weights and central charges to $\pm \infty$), the parent relativistic theory is non-unitary and this is inherited by the non-relativistic theory which, for generic values of certain parameters (non-relativistic conformal weights and central charges), is non-unitary. For specific values of these parameters, there is a sub-sector of the theory that does become unitary, but this sector is also somewhat uninteresting. One of the general features of the non-relativistic theory is that its correlation functions exhibit unfamiliar exponential behaviour in both space and time \cite{Bagchi:2009ca, GCA2d}. When one restricts attention to the unitary theory described above, these exponential pieces vanish. The theory loses all spatial dependence and the correlation functions become trivial and mirror those of a one dimensional relativistic conformal theory.
Given the fact that we have a perfectly fine non-relativistic infinite algebra in two dimensions, it is natural to ask if there are other representations of the algebra which don't suffer from these above difficulties. The answer turns out to be positive. We show that there indeed is a different linear combination of the relativistic Virasoro that also gives the GCA. The earlier contraction of spacetime motivated by non-relativistic scaling however does not seem to make sense here. Interestingly, there is an alternative contraction that is sensible. In this new representation and the alternative contraction scheme, we can compute correlation functions and ask questions about unitarity. One again finds that there are non-trivial pieces to the correlation functions on the one hand and sectors where the non-relativistic theory is unitary. The nice feature now is that imposing unitarity actually leads us to a theory with non-trivial correlation functions.

The representation that is mentioned above and studied in detail through out the paper is natural from the point of view of the BMS/GCA correspondence. One can start of with a bulk $AdS_3$ which has equal left and right moving central charges and generate the BMS through the limit of the radius going to infinity. This representation naturally generates the central charges which can also obtained by an independent analysis in the bulk. The fact that the previous non-relativistic spacetime contraction did not make sense in for this representation was one of the unresolved issues in \cite{Bagchi:2010zz}. With the different contraction, in this paper we resolve this issue.

\subsection{An aside: Non-relativistic Hydrodynamics}
In \cite{Bagchi:2009my}, it was found that the finite GCA is the symmetry algebra of the non-relativistic Euler equations. Interestingly, it was also discovered that a sub-set of the generators of the infinite dimensional GCA turned out to be symmetries of both the Euler and Navier-Stokes equations. This was a re-discovery of the symmetries originally discussed in \cite{Russian}. In fact, as mentioned in the paper where we proposed the BMS/GCA relation \cite{Bagchi:2010zz}, the part of the infinite GCA in two dimensions which is the analogue of the restricted BMS$_3$ algebra (global part of the conformal sub-algebra and the infinite supertranslations) is exactly the symmetry of the Euler equations. So, the symmetries of two-dimensional non-relativistic hydrodynamics is encoded in the asymptotic symmetries of flat space in three dimensions. This is reminiscent of the recent efforts of mapping fluid solutions to flat space geometries \cite{Bredberg:2011jq, Compere:2011dx}. 

As we will comment briefly later, there is more to this surprising connection. Turbulence has been a long sought after problem in theoretical physics which has had very little analytic control. We comment later how 2d GCA is related to simple models of turbulence. Thus studying interesting unitary representations of the 2d GCA can lead us to some possible insight into simplified models of turbulence in two dimensions. This, by the BMS/GCA correspondence is again related to symmetries of three dimensional flat spaces.

\subsection{Outline of the paper}

The outline of the paper is as follows: in Sec 2 we review our understanding of the GCA focussing on the 2d GCA. In Sec.~3, we revisit the BMS/GCA correspondence. In Sec.~4, we introduce the representation of the 2d GCA that we would be working with and show how it can be obtained as a space-time contraction. We go on to discuss the quantum numbers which label the representation and construct the correlation functions. We perform an analysis of the null vectors to understand unitarity issues and discuss the connections to non-relativistic hydrodynamics. In Sec.~5, the bulk interpretation of this new contraction is explained first in terms of Killing vectors in AdS$_3$ and then from the point of view of the asymptotic symmetries. We conclude in Sec.~6 with comments on the contraction in general dimensions and discuss future directions of work. Appendix A contains details of the higher dimensional contraction while Appendix B gives the 4d bulk analysis constructed in parallel to that outlined in Sec.~5 for the 3d case. 

\bigskip

{\em{Note Added:}} When this paper was being readied for submission, we became aware of the upcoming publication \cite{Barnich:adslimt}. Although there is no direct overlap with our work here, both the projects are aimed in the same direction, constructing and understanding the flat space limit of AdS. We are thankful to Glenn Barnich for sharing results with us prior to publication and we believe that their bulk construction would be very useful in understanding the BMS/GCA correspondence further.

\section{A Review of the GCA}


Various non-relativistic version of AdS/CFT have been recently studied in great detail in view of possible connections to strongly coupled condensed matter systems which are inherently non-relativistic \cite{Hartnoll:2009sz, Hartnoll:2011fn, McGreevy:2009xe}. The symmetries of the non-relativistic CFTs on the boundary have been different in these different versions. The most studied of these version has been holography in relation to the Schrodinger algebra, the symmetry algebra of the free Schrodinger equation. Another well-studied example has been the less symmetric Lifshitz algebra. Spacetimes dual to these systems have been found in \cite{Balasubramanian:2008dm, Son:2008ye} and \cite{Kachru:2008yh} respectively.

There is a very precise procedure, called the In\"{o}n\"{u}-Wigner contraction, by which one can obtain the Galilean algebra from the Poincare algebra in any dimensions. The most natural route of a non-relativistic limit of AdS/CFT is to follow this procedure in the relativistic conformal algebra on the boundary and look at its consequences in the bulk dual. This was attempted in \cite{Bagchi:2009my}. The Galilean Conformal Algebra is the finite dimensional algebra that emerges as the limit of the relativistic conformal algebra in this non-relativistic limit. One of the striking features of the GCA is that it can have infinite extensions in any spacetime dimensions{\footnote{The infinite algebra is also called the GCA. To distinguish between the two, we shall explicitly refer to one as the finite GCA and the other as the infinite GCA.}} \cite{Bagchi:2009my}. It has been shown that the maximal set of conformal isometries of Galilean spacetime generates this infinite GCA \cite{Duval:2009vt}. Algebraically, the set of vector fields generating these symmetries are given by
\ben{gcavec}
\tL^{(n)} &=& -(n+1)t^nx_i\p_i -t^{n+1}\p_t \,,\cr
\tM_i^{(n)} &=& t^{n+1}\p_i\,, \cr
\tJ_a^{(n)} \equiv J_{ij}^{(n)} &= & -t^n(x_i\p_j-x_j\p_i)\,,
\een
for integer values of $n$. Here $i=1\ldots (d-1)$ range over the spatial directions and $a$ is an adjoint index of the rotation group. These vector fields obey the algebra
\ben{vkmalg}
[\tL^{(m)}, \tL^{(n)}] &=& (m-n)\tL^{(m+n)}, \qquad [\tL^{(m)}, \tJ_{a}^{(n)}] = -n \tJ_{a}^{(m+n)}, \cr
[\tJ_a^{(n)}, \tJ_b^{(m)}]&=& f_{abc}\tJ_c^{(n+m)}, \qquad  [\tL^{(m)}, \tM_i^{(n)}] =(m-n)\tM_i^{(m+n)}.
\een
The finite dimensional subalgebra consists of taking $n=0,\pm1$ for the
$\tL^{(n)}, \tM_i^{(n)}$ together with $\tJ_a^{(0)}$ and this is the one obtained
by considering the nonrelativistic  contraction of global conformal algebra $SO(d,2)$.
Some subsequent important developments regarding the GCA can be found in \cite{Bagchi:2009ca, GCA2d, Duval:2009vt, alishahiha, RefGCA}. 


In two spacetime dimensions, as is well known, the situation is special.
The relativistic conformal algebra is infinite dimensional and consists of two copies of the Virasoro algebra.
One expects this to be related to the infinite dimensional GCA algebra \cite{GCA2d}. Indeed in two dimensions the
non-trivial generators in  \eq{vkmalg} are the $\tL_n$ and the $\tM_n$:
\be{gca2dvec}
\tL_n = -(n+1)t^n x\p_x -t^{n+1}\p_t\,, \quad \tM_n = t^{n+1}\p_x\,,
\ee
which obey
\be{vkmalg2d}
[\tL_m, \tL_n] = (m-n)\tL_{m+n}\,, \quad [\tM_{m}, \tM_{n}] =0\,, \quad [\tL_{m}, \tM_{n}] = (m-n)\tM_{m+n}
\ee

These generators  in \eq{gca2dvec} arise precisely from a nonrelativistic contraction of the two copies of the Virasoro algebra ($\L_n, \bL_n$). For a small parameter $\e$, at the level of the algebra, we can see if we make the following identifications
\be{Vir2GCA}
\L_n + \bL_n \longrightarrow \tL_{n}\,, \quad \e (\L_n - \bL_n) \longrightarrow - \tM_{n}\,.
\ee
the GCA is generated from the Virasoro algebras, on taking the $\e \to 0$ limit. From a spacetime point of view, we can look at this as a non-relativistic contraction:
\be{nrelscal}
t \rightarrow t\,, \qquad   x \rightarrow \epsilon x\,,
\ee
with $\epsilon \rightarrow 0$. This is equivalent to taking the velocities $v \sim \epsilon$ to zero in units where $c=1$. By looking at the vector fields which generate the centre-less Virasoro Algebra in two dimensions, it is straight forward to check that one obtains \refb{gca2dvec} in the non-relativistic limit. At the quantum level the two copies of the Virasoro get respective central extensions $c, {\bar{c}}$. Considering the linear combinations which give rise to the GCA generators as in \eq{Vir2GCA}, we find
\ben{gcawc}
[\tL_{m}, \tL_{n}] &=& (m-n) \tL_{m+n} + C_1 m(m^2-1) \delta_{m+n,0}\,, \crcr
[\tL_{m}, \tM_{n}] &=& (m-n) \tM_{m+n} + C_2 m(m^2-1) \delta_{m+n,0}\,, \crcr
[\tM_{m}, \tM_{n}] &=& 0\,.
\een
This is the centrally extended GCA in 2d. Note that the relation between central charges is
\be{centch}
C_1 = {{c+\bar c} \over 12}\,, \qquad {C_2 \over \e} = {{\bar c-c} \over 12}\,.
\ee
Requiring a non-zero $C_2$ with a finite $C_1$ can only be achieved if $c$ and $\bar c$ are large (in the limit $\e\rightarrow 0$) and opposite in sign, implying that the original 2d CFT on which we take the non-relativistic limit cannot be generically
unitary. A more detailed analysis of the null vectors of the representation carried out in \cite{GCA2d} tells us that although there exists unitary sectors of the theory, they are relatively less attractive from the point of the correlation functions that we briefly describe next.

The representations of the 2d GCA are constructed by considering the states having definite scaling dimensions \cite{Bagchi:2009ca, GCA2d}:
\begin{equation}
\tL_0 |\tD \rangle = \tD | \tD \rangle \,.
\label{L0=Delta}
\end{equation}
From the algebra, it is clear that $\tL_{n}, \tM_{n}$ with $n >0$ lower the value of the scaling
dimension, while those with $n<0$ raise it. Demanding the
dimension of states be bounded from below primary states in the theory are defined as:
\begin{equation}
\tL_n|\tD \rangle =0\,, \quad
\tM_n|\tD \rangle =0\,,
\label{primop}
\end{equation}
for all $n>0$.
The conditions (\ref{primop})
are compatible with $\tM_0$ in the sense
\begin{equation}
\tL_n \tM_0 |\tD \rangle = 0\,, \quad
\tM_n \tM_0 |\tD \rangle = 0\,,
\end{equation}
and also since $\tL_0$ and $\tM_0$ commute,
an additional label called ``rapidity'' $\tx$ is introduced:
\begin{equation}
\tM_0 |\tD, \tx \rangle = \tx |\tD, \tx \rangle\,.
\end{equation}

Starting with a primary state $|\tD,\tx \rangle$,
one can build up a tower of operators by the action of
$\tL_{-n}$ and $\tM_{-n}$ with $n>0$. The above construction is quite analogous to that of the
relativistic 2d CFT. In fact, from the viewpoint of the limit (\ref{Vir2GCA})
we see that the two labels $\tD$ and $\tx$ are related
to the conformal weights in the 2d CFT as
\be{delxi}
\tD=\lim_{\epsilon \to 0}
(h+\bar h)\,, \qquad  \tx= \lim_{\e\to 0} \e ({\bar{h} -h})\,,
\ee
where $h$ and $\bar h$ are the eigenvalues of $\L_0$ and $\bL_0$, respectively.


The constraints from the Ward identities for the global transformations $\tL_{0,\pm 1}, \tM_{0, \pm 1}$
apply to primary GCA operators and constrain the form of the two point function of primary operators. For primary operators $\O_1(t_1, x_1)$ and $\O_2(t_2, x_2)$ of conformal and rapidity weights $(\tD_1, \tx_1)$ and $(\tD_2, \tx_2)$, the two point function is of the form \cite{Bagchi:2009ca}:
\be{2ptgca}
G_{\rm GCA}^{(2)}(\{t_i, x_i \})
= C_{12} \delta_{\tD_1,\tD_2}
\delta_{\tx_1, \tx_2} t_{12}^{-2\tD_1}
\exp\left( {2\tx_1 x_{12}\over t_{12}} \right).
\ee
Here $C_{12}$ is an arbitrary constant. One can similarly construct the three point function which again is fixed by the symmetries upto an overall constant.

The GCA two and three point functions can also be obtained by taking an appropriate scaling  limit of the usual 2d CFT answers \cite{GCA2d}. This limit requires scaling the quantum numbers of the operators as (\ref{delxi}), along with the non-relativistic limit for the coordinates \eq{nrelscal}.

Let us remind the reader of the issue of unitarity here. In the null vector analysis of \cite{GCA2d}, it was found that there exists a sector of the 2d GCA where the GCA module reduced to the Virasoro module and hence had unitary representation familiar in the study of relativistic 2d CFTs. But this sector required $\xi =0$. All spatial dependence of correlation functions come coupled with $\xi$ dependence so as to survive the contraction limit. Setting $\xi=0$ thus gets rid of all spatial dependence and the correlation functions become ultra-local depending only on time.

\section{The BMS/GCA correspondence}

The computations of the asymptotic symmetry algebra and the central charges in asymtotically AdS$_3$ were known from the classical computations of Brown and Henneaux long before the advent of AdS/CFT and can be seen as one of the earliest motivations behind the Maldacena conjecture. These computations were done at spatial infinity for AdS$_3$.

From the point of view of flat spacetimes, the appropriate boundary from a conformal perspective, is null infinity. If one performs an analysis of the asymptotic symmetry algebra in this context for arbitrary dimensional spacetimes, by solving as in \cite{bar-comp}, the Killing equations to leading order, one ends up with generalisations of the Bondi-Metzner-Sachs algebra. In four dimensions, by not requiring the conformal transformations on the sphere to be well defined globally, one can obtain an enlargement of the symmetry beyond what was originally proposed in \cite{BMS}.

Similar enhancements are possible in three dimensions by relaxing global invariance of the conformal transformations on $S^1$ and now the asymptotic symmetries are given by
\be{}
[ J_m, J_n] = (m-n) J_{m+n}, \quad [J_m, P_n] = (m-n) P_{m+n}, \quad [P_m, P_n] =0.
\ee
One can compute the classical central charges by standard procedures \cite{bar-comp} and the result is
\ben{bms3}
[ J_m, J_n] &=& (m-n) J_{m+n},  \quad [P_m, P_n] =0. \cr
[J_m, P_n] &=& (m-n) P_{m+n} + \frac{1}{4G} m(m^2 -1) \delta_{m+n,0},
\een
These symmetries could be obtained by a contraction of the $AdS_3$ symmetries if one considered \cite{bar-comp}:
\be{bms-gen}
J_n = \L_n - \bL_{-n}, \quad P_n =\frac{1}{\ell} (\L_n + \bL_{-n}).
\ee
Here $\ell$ is the $AdS_3$ radius and this is the parameter that one takes to infinity to obtain flat space.
The $\L$ and $\bL$ are the generators of the asymptotic isometries of AdS$_3$ and their algebra is the Virasoro algebra:
\ben{}
[\L_m, \L_n] &=& (m-n) \L_{m+n} + {c \over 12} m(m^2-1) \delta_{m+n,0}\,,
\nonumber \\[1mm]
[\bL_m, \bL_n] &=& (m-n) \bL_{m+n}
+ {\bar c \over 12} m(m^2-1)\delta_{m+n,0}\,.
\een
The Brown-Henneaux central charges are
\be{}
c = \bar c = \frac{3 \ell}{2G}.
\ee
So, the contraction precisely generates the BMS algebra with the central charges in 3 dimensions.


The BMS/GCA correspondence \cite{Bagchi:2010zz} in three dimensions is the startling observation that \refb{gcawc} and \refb{bms3} are actually the same algebra with $C_1 =0$ and $C_2 = \frac{1}{4G}$.

In four dimensions, the BMS algebra, as stated above, is infinite dimensional and not requiring the generators  to be globally well-defined leads to the following algebra.
\ben{bms4}
[l_m,l_n]=(m-n)l_{m+n},\quad [\bar l_m,\bar l_n]=(m-n)\bar l_{m+n},\quad [l_m,\bar l_n]=0, \crcr
[l_l,T_{m,n}]=(\frac{l+1}{2}-m)T_{m+l,n},
\quad [\bar l_l,T_{m,n}]= (\frac{l+1}{2}-n)T_{m,n+l}.
\een
Now that we have two copies of a centre-less Virasoro algebra, we need to go beyond the usual GCA and the algebra that is isomorphic to this is actually the "semi"-GCA, an algebra which is generated by considering a limit of the 3d-relativistic conformal algebra where only one direction is contracted \cite{alishahiha}. This can be shown to have an infinite extension and is exactly the same as \refb{bms4}.

It was proposed that the correspondence holds for general dimensions. The BMS algebras in higher dimensions are known to reduce to the usual Poincare algebras \cite{Hollands:2003ie, Tanabe:2009va}. These were conjecturally isomorphic to a class of "semi"-GCAs which existed in one lower dimension and were obtained from limits of the corresponding relativistic conformal algebras with only one direction contracted. A quick check tells us the dimensions of the groups match. The Poincare groups are obtained as limits of AdS isometry groups which are isomorphic to the relativistic conformal groups in one lower dimension. The semi-GCA groups are obtained as contractions of the relativistic conformal groups. The process of contraction does not change the number of generators, so the BMS and the "semi"-GCAs have the same dimension.

Now, let us remind the reader of the problems that arose in this surprising relation between flat spaces and non-relativistic symmetries. We will concentrate on the BMS$_3$/GCA$_2$ case. One of the puzzles in trying to push this correspondence further was the hindrance that the non-relativistic contraction \refb{nrelscal} was not compatible with the generators \refb{bms-gen}. On the plane, the contraction made the generators infinite. On the other hand, the other linear combination of the Virasoros \refb{Vir2GCA} which lead to the GCA would require us to start from a non-unitary relativistic theory with infinite positive $c$ and infinite negative $\bar {c}$ if one wanted to generate a non-zero $C_2$ as was required by the BMS algebra and would also require a fine tuning to get $C_1=0$.

We look to resolve this potential tension by defining a new contraction of spacetime which generates the GCA from the Virasoro algebra. This is the contraction that is suited to \refb{bms-gen}. Another interesting thing to note is that we would be doing our computations on the cylinder as opposed to the plane and the contraction seems to be well-defined only here. We remind the reader that the conformal boundary of flat space at null infinity is also a cylinder, in the sense that the spatial sections of the conformal boundary are circles. So there may well be a deeper connection here than meets the eye.

\section{GCA in 2d Redux}

\subsection{GCA in 2d: Different Combination, Different Contraction.}
Motivated by the linear combination of the Virasoro generators that yield the BMS algebra in the limit, in this section we launch a full-scale investigation into this representation of the GCA. We shall first show how one obtains this linear combination from the point of view of a spacetime contraction. We shall consider the relativistic conformal field theory defined on the cylinder.

We consider the following vector fields
\be{}
\L_n = i e^{in\w}\p_{\w}, \quad  \bL_n =  i e^{i n\bw}\p_{\bw}
\ee
where $z=e^{i \w}$ and $\w = t + x$, $\bw = t - x$.
These vector fields generate the centre-less Virasoro algebra. Let us define
\be{repn-redux}
L_n =\L_n - \bL_{-n}, \quad M_n =\e (\L_n + \bL_{-n}).
\ee
So, in the cylinder co-ordinates, the generators take the following form:
\be{} 
\L_n - \bL_{-n} = i e^{inx} \left(i\sin{nt}\, \p_t +  \cos{nt}\, \p_x \right), 
\quad \L_n + \bL_{-n} = i e^{inx} \left( \cos{nt}\, \p_t + i \sin{nt} \,\p_x \right) 
\ee 
Now we define a new contraction by taking the scaling 
\be{newscal} 
t \to \e t, x \to x. 
\ee 
The new vector fields $(L_n, M_n)$ are
well-defined in the $\e \to 0$ limit and are given by 
\be{} 
L_n = ie^{inx}( \p_x+ int \p_t), \quad M_n =  i  e^{inx} \p_t. 
\ee

With the relativistic central charges $c, \bar{c}$ included in the Virasoro algebra, this leads to the 2d GCA
\ben{2dgca}
[ L_m, L_n] &=& (m-n) L_{m+n} + C_1 m(m^2 -1) \delta_{m+n,0},  \quad [M_m, M_n] =0. \cr
[L_m, M_n] &=& (m-n) M_{m+n} + C_2 m(m^2 -1) \delta_{m+n,0},
\een
Here the central charges have the expression $C_1 = \frac{c - \bar{c}}{12}$, $C_2 = \frac{\e ( c + \bar{c} )}{12}$. So, for the case where the relativistic CFT has $c = \bar{c}$, $C_1=0$.

Let us spend a while trying to understand what this new contraction really means. For a Euclidean conformal field theory, where space and time are on the same footing, this is simply a non-relativistic limit like the previous one. Now, for a theory defined on a Lorentzian manifold, this is more subtle. Here let us remember that in dimensions other than two if we took the non-relativistic contraction of $x_i \to \e x_i, t \to t$, we would end up with the finite GCA which would then be lifted to the full infinite GCA. The two dimensional quantum version of the infinite GCA is \refb{gcawc}. We will look at this as the natural definition of a non-relativistic conformal field theory in two dimensions and this new contraction would be essentially a tool to study the representation \refb{repn-redux} from the relativistic infinite-dimensional Virasoro algebra. In some sense, even naively, the new contraction is again non-relativistic. It focusses on a particular spatial slice at an instant of time.

\subsection{Representation and Correlation function}

Now in this representation of the GCA, the states would be labelled again by the eigenvalues under $L_0$ and $M_0$. But we should keep in mind that the role of the two have been interchanged and $M_0$ now labels dilatations and $L_0$ is $\L_0 - \bL_0$.
Accordingly,
\be{}
L_0 | \Delta, \xi \rangle = \xi | \Delta, \xi \rangle , \quad 
M_0 | \Delta, \xi \rangle = \Delta | \Delta, \xi \rangle
\ee
The point to remember here is that
\be{}
\Delta = \lim_{\e \to 0} \e(h + \bar{h}), \quad \xi = \lim_{\e \to 0} (h - \bar{h}).
\ee
Let us now focus our attention on the correlation functions in this representation of the GCA. We will obtain the two-point function as a limit of the relativistic two-point function. The calculations would be done on the cylinder labelled by $(\w, \bw)$.
The correlation function on the plane is given by the familiar expression:
\be{}
G^{(2)} = C z_{12}^{-2h} \z_{12}^{-2 \bar h}
\ee
Now we know, for primary fields $\phi_i$ with weights $h_i, \bar {h}_i$, the correlation functions transform like
\be{}
\< \phi_1(\w_1, \bw_1) \phi(\w_2, \bw_2) \> = \(\frac{dw}{dz}\)^{-h_1}_{\w = \w_1} \(\frac{d\w}{d\z}\)^{-\bar{h}_1}_{\bw = \bw_1}
\(\frac{dw}{dz}\)^{-h_2}_{\w = \w_2} \(\frac{d\w}{d\z}\)^{-\bar{h}_2}_{\bw = \bw_2} \< \phi_1(z_1, \z_1) \phi(z_2, \z_2) \> \nonumber
\ee
Hence the relativistic 2-point function on the cylinder is given by 
\be{2pt-cyl} G^{(2)} = C \(e^{i\w_{12}} (1- e^{-i\w_{12}})^2 \)^{-h}
{\(e^{i\bw_{12}} (1- e^{-i\bw_{12}})^2\)}^{-\bar{h}} 
\ee 
We abuse our notation of the constant $C$ by absorbing some phase factors in it. We will also be simplifying our answer by putting $x_2=0, t_2=0$. 
In order to take the limit we write the above in terms of $(t,x)$.
\ben{}
G^{(2)} &=& C \( e^{i(t+x)} (1 - e^{-i(t+x)})^2\)^{-h} \( e^{i(t-x)} ( 1- e^{-i(t-x)})^2 \)^{- \bar h} \cr 
&=& C \( e^{ix} (1 - e^{-ix})^2 \)^{-h - \bar h} \bigg[
1 +i \(\frac{1+e^{ix}}{1-e^{ix}} \) (h-\bar h) t - \frac{1}{2}
\(\frac{1+e^{ix}}{1-e^{ix}}\)^2 (h-\bar h)^2 t^2 - \cr 
&&
\frac{1}{2} \(\frac{ 2 e^{ix} (h + \bar h) t^2}{(1-e^{ix})^2} \) -i
\frac{1}{6} \(\frac{1+e^{ix}}{1-e^{ix}}\)^3 (h-\bar h)^3 t^3 -i
\frac{e^{ix} (1+ e^{ix}) (h-\bar h) (1+ 3(h+ \bar h)) t^3}{3
(1-e^{ix})^3} + \ldots \bigg] \nonumber 
\een

At this stage, if we naively perform the contraction described above scaling the dilatation and rapidity eigenvalues in the process, it seems we would be left with a null result. The correlation function would be zero {\footnote{except at points where $x_1- x_2= 0 \mod{\pi}$ where we have poles.}}.
To extract a non-zero correlation, we need to focus on certain specific sectors of the GCA. In the non-relativistic algebra, we are free to look wherever we want and we will abuse our freedom by looking near the zero dilatation eigenvalue. We would choose to scale
\be{}
\Delta = {\hat{\e}} \hat{\Delta}
\ee
so that we are looking in the vicinity of the zero eigenvalue in the non-relativistic theory.  This limit, taken simultaneously with the contraction generates the following non-zero correlator{\footnote{We would demand that $\e, \hat{\e}$ go to zero at the same rate.}}:
\be{corr-fn}
G^{(2)} = \tilde{C} (1- \cos{x})^{-\hat \D}
\ee
Here we have absorbed some numerical factors into $\tilde{C}$. We can be even more adventurous and look for more non-trivial pieces. Let us choose to focus on the large-spin sector of the theory
\be{}
\xi = \frac {\hat{\xi}}{\hat{\e}}
\ee
If we do this and simultaneously focus on the near zero dilatation eigenvalue, in this limit, we get the following correlation function:
\be{hs-corr-fn}
G^{(2)} = \tilde{C} (1- \cos{x})^{-\hat \D}\exp\({  \hat{\xi} t \cot(\frac{x}{2})}\)
\ee
A note here about the appearance of the exponential term from the calculation below \refb{2pt-cyl}: the extra terms in the square brackets always scaled at most as $\e^{\alpha-1}$ for a term which had $t^\alpha$ and hence in the contraction always went to zero to leave the exponential piece.

We should remind the reader of the essential difference between this correlation function and the one obtained by the different  contraction on the plane. Here it is $h + {\bar h} = \frac{\D}{\e}$ which is required to be large to survive the contraction for any non-zero $\D$ and the additional exponential piece arises in the limit of large spins which is not required by the theory. On the other hand, in the other representation, we required the spins to be large to survive the contraction and $h + {\bar h}$ did not scale.

Let us note here that if we take any non-zero value of $\Delta$, the above correlation function vanishes (except at points where the denominator vanishes and here we have poles). Hence we focus in the vicinity of $\Delta =0$ by taking the $\hat \e \to 0$ limit. We shall see shortly that unitarity forces us on this $\D=0$ subspace.

\subsection{Unitarity and Null Vectors}

Now we explore the possible unitary sub-sectors of this representation.
The norm of the states in the first level is given by
\be{}
{\mathcal{M}} = \begin{pmatrix} \langle\Delta, \xi|L_1 L_{-1}|\Delta, \xi \rangle & \langle\Delta,
\xi|M_1 L_{-1}|\Delta, \xi \rangle \\ \langle\Delta, \xi|L_1 M_{-1}|\Delta, \xi \rangle & \langle\Delta,
\xi|M_1 M_{-1}|\Delta, \xi \rangle \end{pmatrix} = \begin{pmatrix} 2 \xi & \Delta \\ \Delta & 0 \end{pmatrix}
\ee
which implies
\be{}
\det {\mathcal{M}} = - \Delta^2.
\ee

So, there are explicit negative norm states in the theory with arbitrary $\Delta$. But one has, at least to first level, a unitary truncation if one considers states with $\Delta=0$.

Let us turn our attention back to the correlation functions. First we can look near $\D = 0$ sector which would be unitary in the limit $\hat \e \to 0$. This gives rise to correlators of the form \refb{corr-fn}. Very interestingly, two-point correlation functions of this form have been obtained for scalar operators from a very different perspective in flat space holography recently in \cite{taka} and previously in \cite{Solodukhin:1998ec}. It is fascinating to note the similarity of our results given the very different derivations {\footnote{\cite{taka} also mentions that renormalisation of the apparent singular contribution from the constant in their analysis leads to a vanishing correlation function. We perform our analysis in the boundary non-relativistic conformal theory and do not have anything to say about the constant which multiplies our answer.}}. \cite{taka} also has similar vanishing of correlation functions when one looks at massive scalars, which here would be related to moving away from $\D=0$. We would look to build on this connection between the two apparently different approaches to flat-space holography in the future.

Returning to the present problem, if we focus now on the sector of high spins while remaining in the $\D=0$ eigenspace, then the two-point functions show even more structure and are given by \refb{hs-corr-fn}. From the perspective of the GCA, this is novel because in the previous combination of the Virasoros which generated the GCA, focussing on the unitary sector just left us with time dependent correlation functions.

Even in the sector which has $\D=0$, there would be null vectors in the GCA modules which are built on primary states $| \D, \xi \>$ which needs to be eliminated from the spectrum to get a sensible theory. Null vectors in the GCA, like in the usual Virasoro representation theory, are states that occur for special values of $(\D, \xi)$ and are orthogonal to all states of the module including themselves. One can find the null vector at a given level by considering the most general state at that level built out of the raising operators on a primary state and imposing that lowering operators annihilate the state. This leads to conditions relating different coefficients and also relate the central charges $C_1, C_2$ to $(\D, \xi)$. 

At level one, the most general state is built out of a linear combination of $L_{-1} | \D, \xi \>$ and $M_{-1}| \D, \xi \>$ and the only null state occurs for $\D=0$. For the next level, the most general state is given by 
\be{}
| \chi \> = \left( a_1 L_{-2} + a_2 L_{-1}^2 + b_1 L_1 M_1 + d_1 M_{-1}^2 + d_2 M_{-2} \right ) | \D, \xi \>.
\ee
By following the usual preseciption mentioned above, one can find the null states at this level. The different conditions of the existence of null states and the null states themselves are outlined in the table below \refb{GCAnull}.

\begin{table}[htdp]
\begin{center}
\begin{tabular}{||c|c|c|c||}
\hline
$C_2$& $\D$ & Null State $| \chi^{(1)} \>$ & Null State $| \chi^{(2)} \>$ \\
\hline
\hline
non-zero & non-zero & $(M_{-2} - \frac{3}{4 \D} M_{-1}^2) | \D, \xi \> $& $(L_{-2} - \frac{3}{2\D} L_{-1}M_{-1}+ \frac{3(\xi +1)}{4 \D^2} M_{-1}^2 ) | \D, \xi \>$ \\
\hline
non-zero & zero & $M_{-1}^2 | 0, \xi \>$ & $L_{-1}M_{-1} | 0, \xi =-1 \>$ \\
\hline
zero & zero& $M_{-1}^2 | 0, \xi \>$ & $(L_{-1}M_{-1} - \frac{2(\xi +1)}{3} M_{-2} ) | 0, \xi \>$ \\
\hline
zero & non-zero& no null state & no null state. \\
\hline
\hline
\end{tabular}
\end{center}
\caption{GCA Null States at Level 2}
\label{GCAnull}
\end{table}%

The case when both $C_2$ and $\D$ are zero is interesting as in this case, we can actually see following \cite{GCA2d} that the GCA tower reduces to the Virasoro tower and one can have a truncation to the Hilbert space of this Virasoro module which would have unitary representations in the usual sense. Unlike the case discussed in \cite{GCA2d}, this representation of the GCA actually has interesting correlation functions that we discussed above. 

From the point of view of the BMS/GCA correspondence, we need the sector where $C_2$ is non-zero and $\D$ is zero. We see that there are null states in this sector which need to be dealt with to find a consistent unitary truncation. 

\subsection{Non-Relativistic Hydrodynamics, Turbulence and GCA}

One of the physically interesting questions in the study of Galilean Conformal Algebras is what system this non-relativistic conformal algebra is a symmetry of. In \cite{Bagchi:2009my}, it was already noted the subalgebra of the infinite GCA 
generated by the $M^i_n$ form symmetries of the Navier Stokes equation of non-relativistic hydrodynamics. It was also noticed that the finite GCA ($\tilde{L}_{0, \pm 1}, \tilde{M}^i_{0, \pm1}$) along with this infinite set of generators $M^i_n$ is the symmetry of the Euler equations. (For an extended discussion of the symmetries of the Euler equation in this context, the reader is referred to \cite{thesis}.) All this was, in fact, a rediscovery of long forgotten symmetries of hydrodynamics \cite{Russian}. 

In \cite{Bagchi:2010zz} where the original BMS/GCA correspondence was proposed, it was mentioned that this subset of operators in the 2d GCA was the set that was the "restricted" BMS group in three dimensions, $viz.$ the group obtained by requiring the generators to be well defined everywhere. Thus, the symmetries of two dimensional non-relativistic hydrodynamics encode the symmetries of gravitational physics in one higher dimensional Minkowski spacetime. This is reminiscent of the recent version of the fluid-gravity correspondence which relates non-relativistic Navier Stokes equations and Einstein's equations on higher dimensional Ricci-flat manifolds \cite{Bredberg:2011jq, Compere:2011dx}. 

The fact that we have now found unitary representations of the GCA with correlation functions which are non-trivial has interesting applications for the above. The point to note is that for the two point functions calculated here (and for similar calculations of the three point function), we only needed, like similar computations in relativistic CFTs, the global part of the GCA and hence this would be valid for the non-relativistic fluid systems with the "restricted" infinite group. 

The other very intriguing connection we want to mention here is the possible connection to turbulence {\footnote{AB would like to thank Yaron Oz for enlightening discussions on this issue.}}. The study of turbulence has been long hampered by the complexity of the Navier Stokes equation and the difficulty in any analytic progress in Navier Stokes turbulence (for an excellent review on the subject, see \cite{Gawedzki:1996rk}). This lead researchers to simpler models of stochastic evolution equations. A stochastic evolution equation is of the form  
\be{}
\p_t \Phi = G(\Phi) + f
\ee
where $\Phi$ is a function of time and $G(\Phi)$ a functional local in time. $f$ denotes a stationary Gaussian process with zero mean and variance $ \< f(t) f(s) \> = C (t-s)$. 
The passive scalar model is a particularly nice example of a simple stochastic evolution which is randomly forced at long distances and with $G(\Phi)$ not of the gradient type. This model describes the passive advection of a scalar quantity $T(t, x)$ like temperature by a random velocity field and is given by the equation
\be{pass-scal}
\p_t T = - v^i \nabla_i T + \kappa \nabla^i \nabla_i T + f
\ee
Here $\kappa$ is a constant $\geq 0$, which is called the molecular diffusivity. To simplify the problem, one replaces the velocity field by a Gaussian random field. This model was famously studied by Kraichnan and is a prototypic example of a system where one studies the breakdown of Kolmogorov type scalings. 

Our interest here is the limit where the molecular diffusivity becomes zero. It is clear that in this limit, the system is governed by a Euler-like equation and hence would have an invariance under the "restricted"-GCA in all dimensions. Particularly, in two dimensions, like in the case of the Euler equation, this is thus related to the "restricted"-BMS group. We note the connection between flat-space symmetries and those of simple 2d turbulence models via the BMS/GCA correspondence. Again as in the case of the Euler equations, we believe that the unitary representations and correlation functions found in this paper would be important for the study of this simplified model of turbulence. We hope to report on this intriguing issue in the near future.


\section{Lessons for the Bulk}

In this section, we show that the new contraction that we have discussed at length in the previous section also gives us a proper spacetime realisation of the BMS/GCA correspondence. To this end, we start off by considering the Killing vectors of AdS$_3$, in global co-ordinates. We will show that taking the flat space limit of AdS$_3$ amounts to taking a contraction of the boundary algebra which is the same as \refb{newscal}. We will then reproduce the expression for the symmetry generators by considering the asymptotic generators as a limit of the usual Brown-Henneaux analysis.

\subsection{Contraction of Killing Vectors} \label{contr-killing-3}
We start by looking at the Killing vectors of AdS$_3$ in global co-ordinates. The metric of AdS$_3$ in global coordinates is given by
\be{ads3-global}
ds^2=-\left(1+\dfrac{r^2}{\ell^2}\right)d\tau^2+\dfrac{dr^2}{\left(1+\dfrac{r^2}{\ell^2}\right)}+r^2
d\phi^2.
\ee
where $\ell$ is the radius of AdS space. The Killing vectors which generate the $SO(2,2)$ isometry of this geometry are simply written by using the four-dimensional embedding space. If we choose $(-++-)$ for the signature of four-dimensional embedding space, the killing vectors of \eqref{ads3-global} are given as
\ben{generators of ads3}
\nonumber J_{01}&=&\ell\p_\tau, \quad {} J_{23}=-\p_\phi \\
{}\nonumber J_{02}&=&\sqrt{\ell^2+r^2}
\cos\phi\cos{\tau\over\ell}\,\p_r-{r\ell\over\sqrt{\ell^2+r^2}}\cos\phi\sin{\tau\over\ell}\,\p_\tau-{\sqrt{\ell^2+r^2}\over
r}\sin\phi\cos{\tau\over\ell}\,\p_\phi\\
{}\nonumber J_{03}&=&\sqrt{\ell^2+r^2}
\sin\phi\cos{\tau\over\ell}\,\p_r-{r\ell\over\sqrt{\ell^2+r^2}}\sin\phi\sin{\tau\over\ell}\,\p_\tau+{\sqrt{\ell^2+r^2}\over
r}\cos\phi\cos{\tau\over\ell}\,\p_\phi\\
{}\nonumber J_{12}&=&\sqrt{\ell^2+r^2}
\cos\phi\sin{\tau\over\ell}\,\p_r+{r\ell\over\sqrt{\ell^2+r^2}}\cos\phi\cos{\tau\over\ell}\,\p_\tau-{\sqrt{\ell^2+r^2}\over
r}\sin\phi\sin{\tau\over\ell}\,\p_\phi\\
{}\nonumber J_{13}&=&\sqrt{\ell^2+r^2}
\sin\phi\sin{\tau\over\ell}\,\p_r+{r\ell\over\sqrt{\ell^2+r^2}}\sin\phi\cos{\tau\over\ell}\,\p_\tau+{\sqrt{\ell^2+r^2}\over
r}\cos\phi\sin{\tau\over\ell}\,\p_\phi
\een
To find the generators of the BMS algebra in flat space from this, we would first go near the boundary of AdS$_3$ and then take the limit of $\ell$, radius of AdS$_3$, going to infinity. To go near the boundary, we take $\ell/r \to 0$ limit (finite but large $r$ and fixed $\ell$), the above vectors yield 
\ben{}
J_{1}=\ell\p_\tau &,& \quad J_{6}=-\p_\phi \\
J_{2}=r \cos\phi\cos{\tau\over\ell}\,\p_r -\ell\cos\phi\sin{\tau\over\ell}\,\p_\tau-\sin\phi\cos{\tau\over\ell}\,\p_\phi&,& \quad 
J_{3}=r \sin\phi\cos{\tau\over\ell}\,\p_r -\ell\sin\phi\sin{\tau\over\ell}\,\p_\tau+\cos\phi\cos{\tau\over\ell}\,\p_\phi \nonumber \\
J_{4}=r \cos\phi\sin{\tau\over\ell}\,\p_r+ \ell\cos\phi\cos{\tau\over\ell}\,\p_\tau-\sin\phi\sin{\tau\over\ell}\,\p_\phi &,& \quad
J_{5}=r \sin\phi\sin{\tau\over\ell}\,\p_r + \ell\sin\phi\cos{\tau\over\ell}\,\p_\tau+\cos\phi\sin{\tau\over\ell}\,\p_\phi  \nonumber
\een
We see that the radius of AdS ($\ell$) can be absorbed by defining new time coordinate $t$ as
\be{ptime}  
t={\tau\over \ell}.
\ee
The generators then are given by the following 
\ben{}
J_{1}=\p_t &,& \quad J_{6}=-\p_\phi \\
J_{2}=r \cos\phi\cos{t}\,\p_r -\cos\phi\sin{t}\,\p_t-\sin\phi\cos{t}\,\p_\phi &,&  \quad J_{3}=r \sin\phi\cos{t}\,\p_r -\sin\phi\sin{t}\,\p_t+\cos\phi\cos{t}\,\p_\phi \nonumber \\
J_{4}=r \cos\phi\sin{t}\,\p_r+\cos\phi\cos{t}\,\p_t-\sin\phi\sin{t}\,\p_\phi &,& \quad J_{5}=r \sin\phi\cos{t}\,\p_r+\sin\phi\cos{t}\,\p_t+\cos\phi\sin{t}\,\p_\phi \nonumber
\een
In the $r \to \infty$ limit, these are just the relativistic $2d$ global conformal generators on the cylinder. The CFT is impervious to the scaling of the time co-ordinates for any finite radius $\ell$. This is very much like the conformal scaling of the metric in usual AdS/CFT. The interesting thing happens when we take the flat space limit in the bulk. The generators of the CFT perceive this as a contraction of the time direction as can be seen from \refb{ptime}. 
Taking $\ell \to \infty$ at large but finite $r$ yields the generators near the boundary 
\ben{cont-gen}
J_{1}=\p_t && \quad J_{6}= -\p_\phi \cr
J_{2}=r \cos \phi\,\ \p_r -t\cos\phi\,\p_t-\sin\phi\,\p_\phi && \quad J_{3}=r \sin\phi\,\ \p_r -t\sin\phi\,\p_t+\cos\phi\,\p_\phi \cr
J_{4}=\cos\phi\,\p_t && \quad J_{5}= \sin\phi\,\p_t
\een
 Let us define
 \ben{}
  \nonumber L_0&=&-iJ_6,\qquad L_1=J_2+iJ_3,\qquad L_{-1}=-J_2+iJ_3\\
  {} M_0&=&iJ_1,\qquad M_1=-J_5+iJ_4,\qquad M_{-1}=J_5+iJ_4
  \een
 Thus, the global generators can be re-written in the following way: 
  \be{asy-kill-vec} 
  L_n=ie^{in\phi}(\p_\phi+in t\p_t -inr\p_r),\qquad M_n=ie^{in\phi}\p_t
  \ee
  for $n=-1,0,1$. 
As before, we notice that these generators can be defined for all integral values of $n$ and hence form the 3d BMS algebra.   
\be{}
[L_n,L_m]=(n-m)L_{n+m},\qquad [L_n,M_m]=(n-m)M_{n+m},\qquad
[M_n,M_m]=0.
\ee

It is satisfying to see that on the boundary $r\to \infty$ the above reduces to the expression of the generators of the 2d GCA on the cylinder. So we see that a flat-space limit in the bulk $AdS$ in global coordinates induces a contraction in $t$ of the relativistic conformal algebra reducing it to the 2d GCA and giving a clear physical understanding of how the BMS/GCA correspondence works as a limit of AdS$_3$/CFT$_2$.


\begin{figure}
\begin{center}
\includegraphics{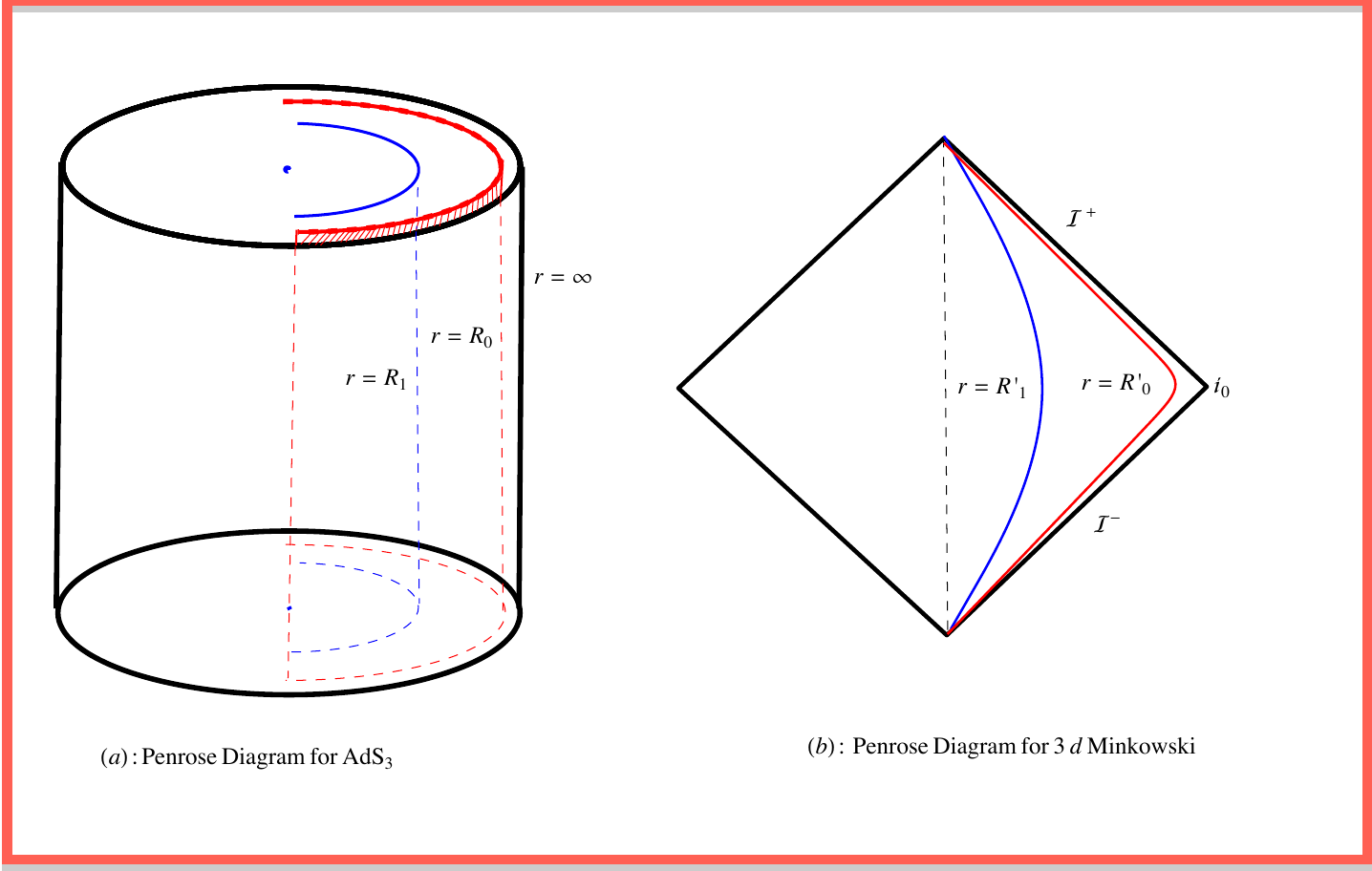}
\caption{Penorse Diagrams for Global $AdS$ and Minkowski spacetimes.}
\end{center}
\end{figure}

An important question to answer is where the CFT lives. In global AdS$_3$, the Penrose diagram is shown above. We indicate some surfaces of constant r in blue and red above. When we take the flat space limit, the same surfaces are depicted by lines of the same colour in the corresponding Penrose diagram in Minkowski spacetime. The relativistic CFT in AdS$_3$ can be thought of as living on those surfaces of constant $r$. The corresponding non-relativistic CFTs would live on the surfaces of constant r in Minkowski spacetime. Taking the limit of $r\to \infty$ in the AdS$_3$ side corresponds to going to the boundary in Minkowski. The boundary in Minkowski are the null boundaries $\mathcal{I}^+, \mathcal{I}^-$ and the point at spatial infinity $i_0$. The 2d GCA which is dual to the flat space at infinity thus lives on all of this boundary. It is interesting to note here that there are indeed structures which are the same as the BMS algebra which exist at spatial infinity $i_0$ \cite{Ashtekar:1978zz}. So this new interpretation does not lead to any inconsistencies. 

The fact that we have been able to generate a representation for the BMS for large but finite $r$ as well as for $r \to \infty$ means that we can extend the matching of symmetries to the near the boundary (red curve) inside Minkowski spacetime
{\footnote{However, we should point out here that this seems to work only in the large $r$ limit i.e. if we choose to take $\ell \to \infty$ for any finite $r$, the algebra of Killing vectors does not form a closed system for any $n$ other than $n=0, \pm1$. There is a notion in terms of a modified bracket \cite{Barnich:bmscft1, Barnich:2011ty, Barnich:2010xq} in which the BMS symmetries can extend into the bulk of Minkowski spacetime. We are currently investigating whether this has some interesting implications for the BMS/GCA correspondence.}}.

 \subsection{Asymptotic symmetry analysis}
 In this subsection, driven by the possibility of extension of the BMS algebra in the bulk, we do a simple-minded application of asymptotic symmetries of Brown and Henneaux by taking the corresponding flat space limit in the bulk of these isometries. We shall be doing our calculations in the global co-ordinates \refb{ads3-global}. The Brown and  Henneaux boundary conditions in these coordinates become
\be{}
\left( \begin{array}{ccc}
   h_{\tau\tau}=\O(1)\, & h_{\tau r}=\O(1/r^3)\, & h_{\tau\phi}=\O(1) \\
      & \,h_{rr}=\O(1/r^4)\,\,\,\, & h_{r\phi}=\O(1/r^3) \\
      &   & h_{\phi\phi}=\O(1) \\
 \end{array}\right).
 \ee
The general form of the asymptotic killing vectors which preserve the above condition has the following expansion
 \ben{}
 \xi^{\tau}_n&=&{\ell\over 2}\,\exp\left(in({\tau\over\ell}\pm
 \phi)\right)+\O(1/r^2)\\
 \xi^{r}_n&=&-{inr\over 2}\,\exp\left(in({\tau\over\ell}\pm
 \phi)\right)+\O(1)\\
 \xi^{\phi}_n&=& \pm {1\over 2}\,\exp\left(in({\tau\over\ell}\pm \phi)\right)+\O(1/r^2)
 \een
The generators of the asymptotic symmetry algebra are given by
\be{}
\L_n^\pm={i\over 2}\exp\left(in({\tau\over\ell}\pm
\phi)\right)\left[ \ell \partial_\tau-inr\partial_r\pm\partial_\phi\right]
\ee
Following the previous analysis, let us define
\be{} L _n=\L_n-\bL_{-n},\qquad M_n=\frac{1}{\ell}(\L_n+\bL_{-n})
\ee
In the limit $\ell\to \infty $ we have
\be{}
L_n= i e^{in\phi}\left(\partial_\phi + i n\tau \partial_\tau - inr\partial_r \right) \quad M_n = i e^{in\phi}\partial_{\tau}.
\ee
It is satisfying to see that we have reproduced the form of the asympototic Killing vectors \refb{asy-kill-vec}. The construction in the previous sections is reminiscent to the initial derivation of the infinite Galilean Conformal Algebra in \cite{Bagchi:2009my} and the one in this section mirrors the reproduction of the GCA in 2d as a limit of the Virasoros \cite{GCA2d}. The emergence of the BMS algebra is obviously not a surprise as we know that the linear combinations were going to reduce to the contracted algebra. What is nice however is the fact that we have re-derived the expressions for the asymptotic Killing vectors which we wrote down by observation in the previous sub-section.

\section{Conclusions and Future Directions}

\subsection{Summary of Results}

In this paper, we have given a concrete spacetime realisation to the mysterious connection between the symmetries of flat space and those arising in non-relativistic conformal systems. After a quick review of the GCA, especially the 2d GCA, and revisiting the statement of the BMS/GCA correspondence, we focussed our attention on a different representation of the 2d GCA. This was obtained at an algebraic level by considering a different linear combination of the two copies of the Virasoro algebra and was realised as a new contraction in which one scaled time and not the spatial direction. We showed that in this representation there were interesting correlation functions which also survived the unitarity constraints. The form of the correlators were intriguingly similar to ones obtained in bulk calculations of flat-space correlation functions in \cite{taka}.

We then turned our attention to the bulk. We showed that the flat-space limit of AdS$_3$ induced the contraction of time on the conformal field theory to give the GCA that we had discussed at length. This was done at the level of the Killing vectors in global AdS$_3$. We also reproduced the expressions for the infinite dimensional asymptotic Killing vectors by taking a combination of the Brown-Henneaux construction. 

\subsection{Various comments}

Let us remind the reader that in the original formulation of the connection, the higher dimensional symmetries were obtained from the CFT perspective by the contraction of a single direction in a d-dimensional CFT. The fact that the time direction is special in Lorentzian spacetimes makes the choice of contracting along time a particularly attractive option. In Appendix~\ref{ApA}, we perform a time-like contraction for general dimensions show that in four dimensions we can indeed generate the infinitely extended "semi"-GCA in this contraction scheme. The corresponding bulk analysis is straight forward: one needs to consider AdS$_4$ in global co-ordinates and the flat space limit there is again a time-like contraction of the generators of the CFT which now give the "semi-GCA". The details of this construction are worked out in Appendix~\ref{ApB}. 

The appearance of the infinite symmetry in the four dimensional case still remains somewhat mysterious. Let us attempt a heuristic argument here. We have seen in the three dimensional case that the flat-space limit is the same as a contraction of the time co-ordinate. Suppose we are close to the boundary and we look at the Poincare co-ordinates of AdS$_4$. 
\be{}
ds^2 = {1 \over{z^2}} \left( dz^2 - dt^2 + dx^2 + dy^2 \right) \to {1 \over{z^2}} \left( dz^2 + dx^2 + dy^2 \right)
\ee
The limit seems to be singling out an (Euclidean) $AdS_3$ factor which is the one principally responsible for the infinite dimensional symmetry algebra. This turns out to be a fibre-bundled $AdS_3 \times R_t$ structure. One can actually show that the BMS$_4$ algebra emerges from this structure of the spacetime as its asymptotic isometries following an analysis which mirrors \cite{alishahiha}. We are still unsure of why this structure should be relevant to flat space physics. 

Returning to the 3d bulk case which we discussed at length, some readers may be worried what lessons this analysis would teach us, given the fact that 3d Minkowski spacetime has no propagating degrees of freedom. There are no non-trivial solutions, not even any asymptotically flat 3d black holes. The absence of dynamical degrees of freedom would potentially show up in the calculation of the correlation functions. It is possible we would find that all the correlation functions of the CFT vanish. We remind the reader that in our analysis we have calculated the correlation functions in the CFT and hence our answers which are dictated by symmetry are valid up to constant factors. By appropriate renormalisation, these factors could very well turn out to be zero in the actual dual theory. 

One could think of adding fields into the bulk and again for appropriate fall offs, one would generate infinite dimensional asymptotic symmetry algebra having currents along with Virasoros factors. Our construction tells us that when we look at the flat space version, we can actually find duals which would be governed not by relativistic CFTs with additional currents, but of contractions of these symmetries. And obviously, to learn more about the physically relevant four dimensional case, one should analyse the lower dimensional case in all details. We are confident that our detailed analysis of the 3d example would teach us valuable lessons for the generalisation to the 4d flat spacetime, where obviously there are dynamical degrees of freedom. 

 \subsection{Future directions}

Coming back to what we have achieved in the present paper, we note that we have been able to find representations of the 2d GCA which are unitary and also contain non-trivial correlation functions. These surprisingly matched with ones from other attempts at flat space holography \cite{taka}. This is something which deserves particular attention and we plan to revisit this issue in the near future. The correlation functions we have found provide a first step to attempts to a formulation of S-matrix theory \cite{Giddings:1999jq, Gary:2009mi} in this context. 
 
We have also stressed that finding unitary representations and correlation functions of the 2d GCA is important for investigations of non-relativistic hydrodynamics, in particular simple models of 2d turbulence. This is another very interesting avenue for further research. 

An obvious question to address is what happens if we consider the BTZ black hole in the bulk instead of the global AdS. The fact that there are no asymptotically flat 3d black holes makes this question interesting. The geometry would naively be ill-defined. We speculate that this might have close connections to similar limits studied in \cite{joan}, especially the second case of the ''pinching orbifold" that the authors study and would correspond to a special nearly massless BTZ near horizon geometry{\footnote{We would like to thank M.~M.~Sheik Jabbari for discussions on this issue.}}. 

The 4d case was addressed only briefly in this work and needs to be understood in much more detail. This is where the power of infinite symmetry would be particularly relevant. The symmetries in AdS$_3$ were already infinite dimensional and this carried over to the flat space in three dimensions. But in AdS$_4$, one does not have infinite symmetries and this only appears in the flat space limit. It would be particularly enriching to consider what these symmetries can teach us about physical observables in four dimensions and understand what we can say about asymptotically flat 4d blackholes using this formalism.

\subsection*{Acknowledgements}
It is a pleasure to acknowledge discussions with Mohsen Alishahiha, Davood Allahbakhshi, Stephane Detournay, Debashis Ghoshal, Rajesh Gopakumar, Daniel Grumiller, James Lucietti, Amir Esmaeil Mosaffa, M. M. Sheik Jabbari, Joan Simon, Andrew Strominger and Marika Taylor. AB would like to thank University of Amsterdam, Utrecht University, the Isaac Newton Institute for Mathematical Sciences (Programme: Mathematics and Applications of Branes in String and M-theory) and Center for the Fundamental Laws of Nature, Harvard University for hospitality at various stages during course of this work. AB is supported primarily by the EPSRC and would also like to acknowledge the support of the Newton Institute, Cambridge and Harvard University.

\section*{Appendices}
\appendix
\section{New Contraction in General dimensions}\label{ApA}
Here we investigate the new contraction of conformal algebra in  the generic d-dimensional space-time. 
Our analysis would be on the plane and different to the case previously discussed. The two-dimensional 
case would be shown to realise the 2d GCA. We however steer clear of the link between the relativistic Virasoro and 
the 2d GCA as that can be defined only on the cylinder for the linear combination we are interested in. 
We show a similar emergence of the 3d "semi"-GCA and again would be confined to the plane. 
This appendix is just an indication of the fact that the general $t$ contraction generates the algebras we want. To make specific connections to the BMS algebras, it is important that we transform the present analysis to the cylinder which we outline for the 4d case in the next appendix. 

Let us start from the generators of conformal algebra in
d-dimensional Minkowski space which consist in rotation
$J_{\mu\nu}$, translations $P_\mu$, dilatation $D$ and special
conformal transformations $K_\mu$:
 \ben{Conformal algebra}
  J_{\mu\nu} &=& -(x_\mu\partial_\nu-x_\nu\partial_\mu) \\
  P_\mu &=& \partial_\mu \\
 D &=& -(x.\partial) \\
  K_\mu &=& -\left(2x_\mu(x.\partial)-(x.x)\partial_\mu\right)
\een We want to define a new contraction of this algebra by taking
the scaling \be{def of contraction} t \to \e\, t,\qquad x_i \to x_i.
\ee In the limit $\e\to 0$, the above generators  reduce to
 \ben{generators }
 \nonumber J_{ij} &=& -(x_i\partial_j-x_j\partial_i),\qquad J_{0i}=x_i\p_t \\
\nonumber P_i &=& \p_i,\qquad\qquad\qquad P_0=\p_t \\
\nonumber D &=& -(t\p_t+x_i\p_i) \\
 K_i &=& -\left[2x_i(x_j\p_j+t\p_t)-(x_jx_j)\p_i\right],\qquad
K_0=x_ix_i\p_t\een which satisfy the following algebra
\ben{contracted algebra1}\nonumber
[J_{ij},J_{rs}] &=& so(d-1),\qquad[J_{ij},J_{0r}]=-(J_{0i}\delta_{jr}-J_{0j}\delta_{ir}) \\
 {} \nonumber [J_{ij},P_{r}] &=& -(P_i\delta_{jr}-P_j\delta_{ir}),\qquad [J_{ij},P_0]=0 \\
 {}\nonumber     [J_{0i},J_{0j}]&=&0,\qquad [P_i,P_j]=0,\qquad[P_0,P_i]=0 \\
 {}[J_{0i},P_j]&=&-\delta_{ij}P_0 ,\qquad[P_0,J_{0i}]=0\een

\ben{contracted algebra2}\nonumber [K_0,K_i]&=&0,\qquad
[K_0,J_{0i}]=0,\qquad
[K_0,P_i]=-2\delta^j_iJ_{0j}\\
{}\nonumber[J_{ij},K_r]&=&K_j\delta_{ir}-K_i\delta_{jr},\qquad
[J_{ij},K_0]=0,\qquad [J_{ij},D]=0\\
{}\nonumber[K_i,K_j]&=&0,\qquad [K_i,J_{0j}]=K_0\delta_{ij},\qquad
[K_i,P_j]=-2D\delta_{ij}-2J_{ij}\\
{}\nonumber[P_0,K_i]&=&-2J_{0i},\qquad [D,K_i]=-K_i,\qquad[D,J_{0i}]=0\\
{}[D,P_i]&=&P_i,\qquad[D,P_0]=P_0,\qquad[P_0,K_0]=0,\qquad[D,K_0]=-K_0
\een
It is clear from the above algebra (specially the last line of
\eqref{contracted algebra1}) that  for the generic d-dimension, this
algebra does not have Galilean sub-algebra and hence it is not GCA.
However, in 2-dimension parameterized by $(t,x)$, one can realize
Galilean subalgebra by identifying
 \be{}B=J_{0x},\qquad P=P_0,\qquad
H=P_x.\ee Moreover, by denoting \ben{}\nonumber L_0&=&D,\qquad
L_1=K_x,\qquad L_{-1}=-P_x\\ M_0&=&J_{0x},\qquad M_1=K_0,\qquad
M_{-1}=P_0 \een we can write the above algebra as
 \be{}
[L_n,L_m]=(n-m)L_{n+m},\qquad [L_n,M_m]=(n-m)M_{n+m},\qquad
[M_n,M_m]=0
 \ee
where $m,n=-1,0,1$.

Using the explicit form \eqref{generators } for the generators of
algebra, we can write \ben{infinite generators}
\nonumber L_n&=&-x^{n}[x\p_x+(n+1)t\p_t]\\
M_n&=&x^{n+1}\p_t\een where $m,n=-1,0,1$. However, we can consider
the generators \eqref{infinite generators} for arbitrary integer
values of $n$ and find an infinite dimensional algebra. This algebra
is isomorphic to the BMS algebra which is the asymptotic symmetry
of 3-dimensional asymptotic flat space-times.

 There is another possibility for finding an infinite extension in
$d=3$. Let us parameterize the coordinates of 3-dimensional
Minkowski space by $\{t,x,y\}$ and define new generators as
\be{}L_0={1\over2}\left(D+iJ_{xy}\right),\qquad
L_{-1}=-{1\over2}\left(P_x+iP_{y}\right),\qquad
L_{1}={1\over2}\left(K_x-iK_{y}\right)\ee \be{}\bar
L_0={1\over2}\left(D-iJ_{xy}\right),\qquad \bar
L_{-1}=-{1\over2}\left(P_x-iP_{y}\right),\qquad \bar
L_{1}={1\over2}\left(K_x+iK_{y}\right)\ee \be{}M_{00}=P_0,\qquad
M_{01}=J_{0x}+iJ_{0y},\qquad M_{10}=J_{0x}-iJ_{0y},\qquad
M_{11}=K_0\ee For the above generators we have \ben{}\nonumber
[L_n,L_m]&=&(n-m)L_{n+m},\qquad \qquad\,\, [\bar L_n,\bar
L_m]=(n-m)\bar L_{n+m}\\ {}
[L_n,M_{rs}]&=&\left({n+1\over2}-r\right)M_{(n+r)s},\quad [\bar
L_n,M_{rs}]=\left({n+1\over2}-s\right)M_{r(n+s)}\een where
$n,m=-1,0,1$. Using \eqref{generators } we find \ben{def of 3d
generators} \nonumber
L_n&=&-{(x-iy)^n\over2}\left[(n+1)t\p_t+(x-iy)(\p_x+i\p_y)\right]\\
{}\nonumber\bar
L_n&=&-{(x+iy)^n\over2}\left[(n+1)t\p_t+(x+iy)(\p_x-i\p_y)\right]\\
M_{nm}&=&(x-iy)^{n}(x+iy)^{m}\p_t\een where $n=-1,0,1$. It is
possible to generalized \eqref{def of 3d generators} for arbitrary
integer $n$ and find an infinite algebra which is isomorphic to the
$BMS$ algebra of 4-dimensional asymptotic flat space-times.

\section{Contraction of AdS$_4$ Isometries and BMS$_4$}\label{ApB}
In this appendix, we give the explicit construction for the contraction of the Killing vectors in $AdS_4$. The story would be analogous to the discussion in Sec.~\refb{contr-killing-3}. For the convenience of the readers interested in the details of the construction, we present our calculation in full detail. 

\noindent 
Let us consider AdS$_4$ space-times with metric
  \be{}
  ds^2=-\(1+{r^2\over\ell^2}\)dt^2+{dr^2\over
  \(1+{r^2\over\ell^2}\)}+r^2d\Omega^2.
  \ee
The generators for the $SO(2,3)$ isometry of this geometry are
 \ben{}
 \nonumber J_1&=&-\cos\phi\,\p_\theta+\cot\theta\sin\phi\,\p_\phi\\
 J_2&=&-\sin\phi\,\p_\theta-\cot\theta\cos\phi\,\p_\phi\\
 \nonumber J_3&=&-\p_\phi\\
 \nonumber J_4&=&-{r\ell\over\sqrt{r^2+\ell^2}}\sin {t\over\ell}\cos\theta\,\p_t+\sqrt{r^2+\ell^2}\cos{t\over\ell}\cos\theta\,\p_r-{\sqrt{r^2+\ell^2}\over r}\cos {t\over\ell}\sin\theta\,\p_\theta\\
 \nonumber J_5&=&-{r\ell\over\sqrt{r^2+\ell^2}}\sin {t\over\ell}\sin\theta\cos\phi\,\p_t+\sqrt{r^2+\ell^2}\cos{t\over\ell}\sin\theta\cos\phi\,\p_r+{\sqrt{r^2+\ell^2}\over r}\cos {t\over\ell} \(\cos\theta\cos\phi\,\p_\theta-{\sin\phi\over\sin\theta}\p_\phi\)\\
  \nonumber J_6&=&-{r\ell\over\sqrt{r^2+\ell^2}}\sin {t\over\ell}\sin\theta\sin\phi\,\p_t+\sqrt{r^2+\ell^2}\cos{t\over\ell}\sin\theta\sin\phi\,\p_r+{\sqrt{r^2+\ell^2}\over r}\cos {t\over\ell}\(\cos\theta\sin\phi\,\p_\theta+{\cos\phi\over\sin\theta}\p_\phi\)\\
  \nonumber J_7&=&-{r\ell\over\sqrt{r^2+\ell^2}}\cos {t\over\ell}\cos\theta\,\p_t-\sqrt{r^2+\ell^2}\sin{t\over\ell}\cos\theta\,\p_r+{\sqrt{r^2+\ell^2}\over r}\sin {t\over\ell}\sin\theta\,\p_\theta\\
   \nonumber J_8&=&-{r\ell\over\sqrt{r^2+\ell^2}}\cos {t\over\ell}\sin\theta\cos\phi\,\p_t-\sqrt{r^2+\ell^2}\sin{t\over\ell}\sin\theta\cos\phi\,\p_r-{\sqrt{r^2+\ell^2}\over r}\sin {t\over\ell}\(\cos\theta\cos\phi\,\p_\theta+{\sin\phi\over\sin\theta}\p_\phi\)\\
  \nonumber J_9&=&-{r\ell\over\sqrt{r^2+\ell^2}}\cos {t\over\ell}\sin\theta\sin\phi\,\p_t-\sqrt{r^2+\ell^2}\sin{t\over\ell}\sin\theta\sin\phi\,\p_r-{\sqrt{r^2+\ell^2}\over r}\sin {t\over\ell}\(\cos\theta\sin\phi\,\p_\theta+{\cos\phi\over\sin\theta}\p_\phi\)\\
  J_{10}&=&\ell\p_t \nonumber
 \een
 Taking the large radius limit with finite $\ell$ we go near the boundary. The Killing vectors in this limit become
 \ben{}
 \nonumber J_1&=&-\cos\phi\,\p_\theta+\cot\theta\sin\phi\,\p_\phi, \quad 
  \nonumber J_2 =-\sin\phi\,\p_\theta-\cot\theta\cos\phi\,\p_\phi\\
  J_3&=&-\p_\phi, \quad J_{10} = \ell\p_t \\
 \nonumber J_4&=&-\ell\sin {t\over\ell}\cos\theta\,\p_t+r\cos{t\over\ell}\cos{\theta}\,\p_r-\cos {t\over\ell}\sin\theta\,\p_\theta\\
 \nonumber J_5&=&-\ell\sin {t\over\ell}\sin\theta\cos\phi\,\p_t+r\cos{t\over\ell}\sin{\theta}\cos\phi\,\p_r+\cos {t\over\ell}\cos\theta\cos\phi\,\p_\theta-{\cos {t\over\ell}\sin\phi\over\sin\theta}\p_\phi\\
  \nonumber J_6&=&-\ell\sin {t\over\ell}\sin\theta\sin\phi\,\p_t+r\cos{t\over\ell}\sin{\theta}\sin\phi\,\p_r+\cos {t\over\ell}\cos\theta\sin\phi\,\p_\theta+{\cos {t\over\ell}\cos\phi\over\sin\theta}\p_\phi\\
  \nonumber J_7&=&-\ell\cos {t\over\ell}\cos\theta\,\p_t-r\sin{t\over\ell}\cos{\theta}\,\p_r+\sin {t\over\ell}\sin\theta\,\p_\theta\\
   \nonumber J_8&=&-\ell\cos {t\over\ell}\sin\theta\cos\phi\,\p_t-r\sin{t\over\ell}\sin{\theta}\cos\phi\,\p_r-\sin {t\over\ell}\cos\theta\cos\phi\,\p_\theta+{\sin {t\over\ell}\sin\phi\over\sin\theta}\p_\phi\\
  \nonumber J_9&=&-\ell\cos {t\over\ell}\sin\theta\sin\phi\,\p_t-r\sin{t\over\ell}\sin{\theta}\sin\phi\,\p_r-\sin {t\over\ell}\cos\theta\sin\phi\,\p_\theta-{\sin {t\over\ell}\cos\phi\over\sin\theta}\p_\phi
 \een
 Now we take the flat-space limit by scaling $\ell\to\infty$. It is not difficult to show that the new generators
 \ben{}
\nonumber L_0&=&-\frac12\(J_4+iJ_3\),\qquad\qquad\qquad\quad \bar L_0=-\frac12\(J_4-iJ_3\)\\
 \nonumber L_1&=&\frac12\left[\(J_5-J_1\)+i\(J_6-J_2\)\right],\qquad \bar L_1=\frac12\left[\(J_5-J_1\)-i\(J_6-J_2\)\right]\\
 L_{-1}&=&-\frac12\left[\(J_5+J_1\)-i\(J_6+J_2\)\right],\qquad \bar L_{-1}=-\frac12\left[\(J_5+J_1\)+i\(J_6+J_2\)\right]
 \een

 \ben{}
 \nonumber M_{0,0}&=&\frac{1}{\ell} \(J_{10}+J_7\),\qquad\quad M_{1,1}=\frac{1}{\ell}\(J_{10}-J_7\)\\
  M_{0,1}&=&\frac{1}{\ell}\(-J_{8}+iJ_9\),\qquad M_{1,0}=\frac{1}{\ell}\(-J_{8}-iJ_9\)
 \een
are well-defined in this limit and satisfy the following algebra 
\ben{}
 \nonumber [L_m,L_n]&=&(m-n)L_{m+n},\qquad [\bar L_m,\bar L_n]=(m-n)\bar L_{m+n}\\
 \nonumber [L_m,\bar L_n]&=&0,\qquad\qquad \qquad\qquad [L_l,M_{m,n}]=\({l+1\over2}-m\)M_{m+l,n}\\
   {[}\bar L_l,M_{m,n}{]}&=& \({l+1\over2}-n\)M_{m,n+l}
 \een
for $m,n=-1,0,1$. Notice that again, as in the 3d case, the flat space limit manifests itself as a contraction of the time direction in the boundary theory. To find nice explicit expressions of our above generators we switch to polar coordinates
 \be{}
 x=e^{i\phi}\cot\frac\theta 2,\qquad y=e^{-i\phi}\cot\frac\theta 2 .
 \ee
With this the above generators can be written as
 \ben{}
 \nonumber L_n&=& \frac12\({xy-1\over xy+1}-n\)x^n\(t\p_t-r\p_r\)-x^{n+1}\p_x\\
 \nonumber \bar L_n&=& \frac12\({xy-1\over xy+1}-n\)y^n\(t\p_t-r\p_r\)-y^{n+1}\p_y\\
 M_{m,n}&=&{2\over 1+xy}x^m y^n \p_t
 \een
 We again observe that the generators can be defined for any integer $n$ and the algebra can be given an infinite dimensional lift. This is the infinitely extended bms$_4$ \cite{Barnich4d, Barnich:bmscft1}. At the boundary, $r \to \infty$, this reduces to the semi-GCA. We note here that our expressions are in complete agreement with corresponding results of \cite{Barnich:bmscft1}.
It would be particularly interesting in the context of the BMS/GCA correspondence to see what role the modified bracket introduced in \cite{Barnich:bmscft1} plays in extending the BMS algebra into the bulk and similarly, what it means for the boundary contracted conformal field theory.



\end{document}